\documentclass[useAMS,usenatbib,usegraphicx]{mn2e}

\newcommand{\vk}{\mbox{\bf {k}}}
\newcommand{\vkt}{\hat{ \mbox{\bf {k}}}}
\newcommand{\vzt}{\hat{ \mbox{\bf {z}}}}
\newcommand{\vq}{\mbox{\bf {q}}}
\newcommand{\vp}{\mbox{\bf {p}}}

\newcommand{\n}{\hat{\mbox{\bf {n}}}}
\newcommand{\np}{\hat{\mbox{\bf {n}}}'}

\newcommand{\pmin}{p_{\mbox{min}}}
\newcommand{\pmax}{p_{\mbox{max}}}

\input epsf

\def\plotancho#1{\includegraphics[width=18cm]{#1}}

\title{On the Influence of Resonant Scattering on Cosmic Microwave Background
Polarisation Anisotropies}

\author[C. Hern\'andez--Monteagudo, J.A. Rubi\~no-Mart\1n \& R.A. Sunyaev]{C.
Hern\'andez--Monteagudo$^{1}$\thanks{E-mail:carloshm@astro.upenn.edu},
J.A. Rubi\~no--Mart\1n$^{2}$ and R.A. Sunyaev$^{3,4}$\\
$^{1}$Department of Physics and Astronomy, University of Pennsylvania, 209 South 33rd Str., 19104-6393 Philadelphia, USA\\
$^{2}$Instituto de Astrof\1sica de Canarias, C/ V\1a L\'actea, s/n, 38200, La Laguna, Tenerife, Spain\\
$^{3}$Max Planck Institut f\"ur Astrophysik (MPA), Karl-Schwarzschild Str. 1, D-85740, Garching bei M\"unchen, Germany\\
$^{4}$Space Research Institute (IKI), Profsoyuznaya 84/32, Moscow 117810, Russia 
}
\begin{document}

\date{}

\pagerange{\pageref{firstpage}--\pageref{lastpage}} \pubyear{2006}

\maketitle

\label{firstpage}

\begin{abstract}
  We implement the theory of resonant scattering in the context of
  Cosmic Microwave Background (CMB) polarisation anisotropies. We
  compute the changes in the E-mode polarisation (EE) and Temperature
  E-mode (TE) CMB power spectra introduced by the scattering on a
  resonant transition with a given optical depth $\tau_X$ and
  polarisation coefficient $E_1$. The latter parameter, accounting for
  how anisotropic the scattering is, depends on the exchange of
  angular momentum in the transition and enables, {\it a priori}, to
  distinguish among different possible resonances. We use this
  formalism in two different scenarios: cosmological recombination and
  cosmological reionisation. In the context of cosmological
  recombination, we compute predictions in frequency and multipole
  space for the change in the TE and EE power spectra introduced by
  scattering on the $H_{\alpha}$ and $P_{\alpha}$ lines of
  Hydrogen. This constitutes a fundamental test for the standard model
  of recombination, and the sensitivity it requires is comparable to
  that needed in measuring the primordial CMB B-mode polarisation
  component. In the context of reionisation, we study the scattering
  off metals and ions produced by the first stars, and find that
  polarisation anisotropies, apart from providing a consistency test
  for intensity measurements, give some insight on how reionisation
  evolved: since they keep memory of how anisotropic the line
  scattering is, they should be able to discern the OI 63.2$\mu$m
  transition from other possible transitions associated to OIII, NII,
  NIII, etc. The amplitude of these signals are, however, between 10
  to 100 times below the (already challenging) level of CMB B-mode
  polarisation anisotropies.

\end{abstract}

\begin{keywords}
cosmic microwave background -- galaxies: clusters: general -- diffuse radiation -- intergalactic medium
\end{keywords}

\section{Introduction}

The cosmic microwave background (CMB) constitutes one of the
most powerful tools in cosmology today.  Observations of the angular power
spectrum of the CMB intensity and polarisation anisotropies are
providing constraints at the level of few percent on the values of the
cosmological parameters describing our Universe within the standard
cosmological scenario, (e.g.  \citet{wmap3}). Moreover, because these
anisotropies are intricately linked to the initial conditions 
in our Universe, they can also be used as a probe
of inflation and fundamental physics. Microwave intensity (or
temperature) anisotropies have recently been measured in the whole sky
from large to medium ($\sim 0.2\degr$) angular scales by the {\it
Wilkinson Microwave Anisotropy Probe}, (WMAP\footnote{WMAP's URL site:
{\tt http://map.gsfc.nasa.gov}}). WMAP measurements have been
complemented by other ground-based high-angular resolution CMB
experiments \citep{vsa,acbar,cbi} with partial sky coverage. \\

On the other hand, the study of the CMB polarisation anisotropies is
just starting: experiments like CAPMAP \citep{capmap} and CBI
\citep{cbipol} announced the first detection of CMB polarisation
anisotropies in subdegree angular scales, whereas WMAP's team managed
to separate large-scale CMB polarisation anisotropies from the
dominant contamining signals produced in our Galaxy. Indeed, when trying
to achieve better constraints on the theoretical models and to test the
inflationary predictions, polarisation measurements are known to play
an important role in breaking some of the degeneracies present in the
parameters \cite[e.g.][]{page06}. Furthermore, a particular type of
CMB polarisation anisotropies having particular symmetry properties
(the so-called $B$ modes of polarisation) are known to be closely
related to tensorial modes in the metric perturbations (that is, {\em
gravitational waves}) generated during the inflationary
epoch. Therefore, it is easy to understand that the search for these
polarisation anisotropies provides a unique test for both the
cosmological model and fundamental physics, since to date no detection
of gravitational waves has been reported.

For this reason, in the last few years an important effort has been
put into measuring the polarisation angular power spectrum of the CMB,
and consequently there is already a considerable number of projects
(Clover, QUIET, CMBPol) pursuing the detection of the primordial B-mode of the
CMB polarisation anisotropies.  The predicted sensitivities of these
experiments ($10^{-2}$ to $10^{-4}$~$(\mu K)^2$) will open the
possibility to investigate additional interesting effects whose amplitudes
were regarded as unreachable.

In particular, the resonant scattering on Hydrogen Balmer and Paschen
lines at the epoch of recombination ($z\sim 1050$) is known to produce
frequency-dependent temperature anisotropies in the sub-degree angular
scales of the CMB, \citep{jalchmras}. These same lines are known to
give rise to spectral distortions in the CMB black body spectrum
\citep{RCS06, CRS06}, at the relative level of $\sim 10^{-8}$, and in
general of the order of $\sim 10^{-7}$ for higher series. Here we
analyse the changes that these transitions introduce in the
polarisation angular spectrum, and make definite predictions for the
standard model of recombination. The amplitudes of the changes in the
polarisation power spectrum are found to be comparable to the typical
amplitudes of the B modes for values of the tensor-to-scalar ratio $r
\sim 10^ {-2} - 10^{-3}$.  These signals should constitute a new probe
for the end of recombination process, and may provide information
about cosmological parameters and their evolution with time.

Another physical process that upcoming CMB experiments may be
sensitive to is the resonant scattering off metals and molecules at
high redsfhift. This idea is not new, and several authors have studied
the effect of molecules (LiH, DH$^+$) on the CMB radiation at high
redshift, both from the theoretical \citep{dub1,dub2} or the
observational point of view, \citep{dB93}.  In \citet{bhms} (hereafter
BHMS), a detailed theoretical study of the effect of resonant
scattering of CMB photons off metals and ions was presented. In that
work, it was shown that different observing frequencies would probe
the resonant scattering at different redshifts. For this reason,
the comparison of CMB maps obtained at different frequencies should
provide information of the metal enrichment history of the
Universe. This is essentially a consequence of the frequency dependent
optical depth ($\tau_{\nu}$) that these resonant transitions give rise
to. In their analysis, BHMS studied fine structure transitions of
atoms and ions (CI, CII, NII, NIII, OI, OIII, etc), and found that the
scattering process should be dominant over collisional
excitation. Further, they exploited the fact that, as shown in
\citet{hms}, the changes in the CMB temperature angular power spectrum
are linear in $\tau_{\nu}$, and proposed a method to extract the metal
induced signal in upcoming CMB data sets. 

\citet{hmvj06} addressed the requirements that future CMB data must
fulfill in order to be sensitive to the metal-induced signal. Among
all of them, the most demanding was the cross-channel calibration,
whose relative uncertainty had to be below $\tau_{\nu}$, i.e., better
than $\sim 10^{-4}$. Point spread function reconstruction errors were
found not to be very critical. The observing channels should be placed
in an optimal frequency range of [100, 350] GHz, enabling an accurate
subtraction of the thermal Sunyaev-Zel'dovich effect, \citep{tSZ}.

In this paper, we study the changes that resonant scattering off
metals introduce in the CMB polarisation angular power spectrum. The
amplitude of these changes is found to be extremely small, still 
a factor 10 -- 100 below the nominal sensitivity of proposed B-mode
experiments. However, they would provide a parallel test for the
enrichment history of the Universe, and help to distinguish the
OI among other likely resonant species, like NII, NIII, CI, CII or
even OIII.

This study should be understood in the context of proposed CMB
polarisation measurements, that attempt to measure primordial B-modes of typical
amplitudes $10^{-3}$ $(\mu K)^2$. Distinguishing signals of such small
amplitudes demands stringent requirements not only on the sensitivity side, but
also (and perhaps more importantly) on the systematic/foreground side:
contaminants are now known to be orders of magnitude larger
that the targeted B-modes, and their exquisite control and removal
will be absolutely demandatory. The signals we are addressing in this
paper have comparable amplitudes to the intrinsic B-modes: if the latter are
finally measured, then these signals, containing a wealth of
information about the recombination and reionisation histories of our
Universe,  may be accessible too.

The structure of the paper is as follows. In section 2 we implement
the theory of resonant scattering in the context of CMB angular
anisotropies. In Section 3 we apply our formalism to the polarisation
changes introduced by the resonant scattering on $H_{\alpha}$ and
$P_{\alpha}$ lines of Hydrogen during cosmological recombination. In
Section 4 we address the resonant scattering on metals and ions during
the end of the Dark Ages and the beginning of reionisation. We discuss
our results and conclude in Section 5.

\section{Resonant Scattering and CMB Polarisation Anisotropies}

\subsection{Background}

We start by reviewing well known
results on resonance radiation and CMB polarisation. These are
combined in the next subsection in order to compute the changes in the
CMB polarisation power spectrum for different types of transitions. The
reader familiar with the formalism introduced here may therefore
prefer to skip this subsection and continue in the next one.

An analysis of the polarisation state of resonance
radiation coming out of a generic atomic system is presented in
\citet{hamilton} and \citet{chandrasekhar}. Their main result
is that the resonant scattering is very close to Rayleigh scattering, although
in the general case it gives rise to a smaller degree of polarisation. Indeed,
the resonant scattering can be decomposed into a Rayleigh-like scattering term
(i.e., with its same angular dependence) and an isotropic scattering term.  If
$I_{\parallel}'$ and $I_{\perp}'$ denote the incident light intensities parallel
and perpendicular to the scattering plane in a solid angle element around
direction $\np $, then the scattered counterparts ($I_{\parallel}$ and
$I_{\perp}$) along direction $\n$ are given by
\[
\left( \begin{array}{c}
I_{\parallel} \\
I_{\perp} \end{array} \right)
\; d\Omega(\n) = 
\; \tau_{X}\biggl[ E_1\; \frac{3}{8\pi}
\left( \begin{array}{cc}
\cos^2{\theta} & 0 \\
0 &   1 \end{array}  \right)
\; + 
\]
\begin{equation}
\phantom{xxxxxxxx}
\; E_2\;\frac{1}{8\pi} \left( \begin{array}{cc}
1 & 1 \\
1 & 1 \end{array} \right) \biggr]\; 
\left( \begin{array}{c}
I_{\parallel}' \\
I_{\perp}' \end{array} \right)
\; d\Omega(\np).
\label{eq:rseq1}
\end{equation}
$\theta$ and $\tau_{X}$ are the scattering angle and optical depth,
respectively.  The coefficients $E_1$ and $E_2$ provide the relative
weight of the Rayleigh-like and isotropic scatterings, and verify the
condition $E_1 + E_2=1$, whereas the other coefficients (proportional
to $1/4\pi$) ensure conservation of the scattered total intensity. The
values of $E_1$ and $E_2$ depend on the initial value and change
of the quantum number describing the angular momenta involved in the
resonant transition, \citep{hamilton}.

We next write the radiative transfer equation for this scattering in
terms of the Stokes parameters I ($\equiv I_{\parallel} + I_{\perp}$)
and Q ($\equiv I_{\parallel} - I_{\perp}$). This scattering produces
no circular polarisation, so the Stokes parameter V will be ignored.
We study the radiation coming from an arbitrary point $\n$ on the
celestial sphere. Our axes used to define the parallel ($\parallel$)
and perpendicular ($\perp$) directions are determined by the meridian
and azymuthal directions at $\n$, respectively. As shown in
\citet{chandrasekhar}, in order to compute the parameters I$(\n)$ and
Q$(\n)$ in this reference frame, it is necessary to integrate in all
possible directions $\np$ of incoming radiation. In this step, one
must account for the different reference frames being
involved. Since our scattering matrices (eq.(\ref{eq:rseq1})) are
defined with respect to the plane of scattering, it is necessary to
make two transformations of the intensities: the first one projects
intensities at $\np$ into the axes parallel and normal to the
scattering plane, whereas the second transforms the scattered
intensities into our reference frame at $\n$.  Under an
infinitesimally small optical depth due to the resonant transition
($\Delta\tau_{X}$), the radiative transfer equation yields the
following changes in the I and Q parameters for the resonant
scattering:
\[
\Delta \left( \begin{array}{c}
  I \\
  Q \end{array}\right) (\n ) = \; \Delta\tau_{X}\; \biggl[ 
 - \left(\begin{array}{c} I \\
       Q \end{array}\right) (\n ) \;\; + \;
\]
\begin{equation}
\phantom{xxxxxxxxxxx}
\left( \begin{array}{c} 
  I_0 \;-\; E_1\; P_2(\mu )\frac{1}{2}\bigl( I_2 + Q_2 + Q_0\bigr) \\
 E_1\; \bigl(1-P_2(\mu )\bigr) \frac{1}{2}\bigl( I_2 + Q_2 + Q_0\bigr) 
  \end{array}\right) \biggr] .
\label{eq:rseq2}
\end{equation}
In this equation, the parameters I and Q have been expanded on a Legendre
polynomial basis ($X(\n ) = \sum_l (-i)^l (2l+1) P_l(\mu ) X_l$, with $X=I,Q$),
$\mu \equiv \cos \theta$ and $\theta$ is the polar angle at $\n$. Both I and Q
are assumed to be azymuthally symmetric.  This expression meets the standard
Rayleigh scattering limit if $E_1 \rightarrow 1$, and shows how the intensity
quadrupole alone ($I_2$) acts as a source for linear polarisation. At the same
time, it is clear that different resonant transitions with different $E_1$
coefficients give rise to different levels of polarisation.  We shall show
below that this fact may have its relevance, since polarisation measurements may
distinguish among different candidate resonant species otherwise undiscernible
with intensity measurements only: the polarisation term $\Pi \equiv
(I_2+Q_2+Q_0)$ is typically a few percent of $I_0$, and this fact largely hides
the value of $E_1$ from intensity observations.

In a cosmological context, we follow \citet{mb} to describe the
evolution of the perturbations of the photon field. They work in
Fourier space, and therefore the photon distribution function
$f_{\gamma}(\vk,\vq,\eta )$ is {\it a priori} dependent on comoving
wavemodes $\vk$, photon momenta $\vq$ and conformal time
$\eta$. However, in cosmological perturbation theory, due to a
symmetry present in the evolution equations, it is customary to work
with the simplifying assumption that $f_{\gamma}$ depends on the
direction of the photon momentum $\vq / q = \n$ via the dot product
$\n\cdot \vkt = \n \cdot \vk / k$ only, ($k$ being the modulus of
$\vk$ and $q$ the modulus of $\vq$).  This permits to expand the
photon distribution function for the polarisation state $i$ as
\[
 f_{\gamma}^i(\vk,\vq,\eta) = f_0^i(q,\eta)\;\biggl[1+\Psi^i(\vk,q, \vkt\cdot\n,\eta)\biggr] =
\]
\begin{equation}
\phantom{xxxxxx}
 f_0^i(q,\eta)\;\biggl[1+\sum_l (-i)^l(2l+1)P_l(\vkt\cdot\n) \Psi^i_l(\vk, q, \eta)
 \biggr],
\label{eq:phdf1}
\end{equation}
with $ f_0^i(q,\eta) = 1/h^3(\exp{(qc/k_BT_{CMB}[\eta ])}-1)$ the
unperturbed Planck distribution function for the polarisation state
$i$. $ \Psi^i_l$ is the $vk$, $q$ and $\eta$ dependent perturbation of
the distribution function for the same polarisation state $i$. This
expansion can be applied to the momentum-averaged energy distribution
function for both polarisation states,
\begin{equation}
F_{\gamma, l}(\vk, \eta) = \frac{\sum_{i=1}^2\int
  dqq^2f_0^i(q,\eta)q\Psi_l^i}{\sum_{i=1}^2\int dqq^2f_0^i(q,\eta)q},
\label{eq:fgammal}
\end{equation}
and for the difference of the two polarisation components,
\begin{equation}
G_{\gamma, l}(\vk, \eta) = 
  \frac{\int dqq^2q\bigl(f_0^{1}(q,\eta)\Psi_l^{1}-f_0^{2}(q,\eta)\Psi_l^{2}\bigr)}{\sum_{i=1}^2\int dqq^2f_0^i(q,\eta)q}.
\label{eq:ggammal}
\end{equation}
These two $l$-dependent quantities correspond to the $X_l$ terms in the
expansion $X(\vk,\vq,\eta) = \sum_l (-i)^l (2l+1) P_l(\vkt\cdot\n) X_l(\vk, q,
\eta)$, with $X=F_{\gamma},G_{\gamma}$.  If now a reference system for $\vq$ is
chosen so that the $\vzt$ axis is parallel to $\vkt$, then it is clear that the
linear polarisation will be in the direction of changing $\theta$, i.e.,
parallel to $\partial / \partial \theta $, due to the azymuthal symmetry of
$f_{\gamma}(\vk, q, \vkt\cdot\n, \eta)$ for fixed $\vk$ and $\eta$.  Therefore,
in the orthonormal basis $\{\partial/\partial\theta, \partial/\partial \phi\}$
the momentum-integrated quantity $G_{\gamma}$ corresponds to the Q Stokes
parameter, whereas U vanishes. \citet{zaldaharari} formally integrated the
evolution equations for $\Delta_T \equiv F_{\gamma} / 4$ and $\Delta_P \equiv
G_{\gamma}/4$ for every Fourier mode $\vk$, (note that we have switched into
temperature units).  However, since the Q and U Stokes parameters are dependent
on the orientation of the reference frame where they are defined, when
integrating in all Fourier modes to obtain $\Delta_T,\Delta_P$ along a given
line of sight $\n$ the following problem arises: for every new $\vk$ a new
(rotated) reference system must be adopted, and Q and U must be transformed
accordingly. \citet{zaldaseljak97} solved this problem by introducing the
combinations $Q(\n)\pm iU(\n)$ that can be easily transformed into {\em
rotationally invariant} quantities $E(\n)$ and $B(\n)$. After projecting the
three quantities $\Delta_T(\n)$, $E(\n)$ and $B(\n)$ on a spherical harmonic
basis,
\begin{equation}
\Delta_{T,E,B} = \sum_{l,m} a_{l,m}^{T,E,B} Y_{l,m}(\n),
\label{eq:alms}
\end{equation}
they found analytical expressions for the angular power spectra
$C_l^{T,E,B} = \langle a_{l,m}^{T,E,B} \left( a_{l,m}^{T,E,B}\right)^*\rangle$:
\begin{equation}
C_l^{T,E}  = 
 (4\pi)^2 \int dk\; k^2\; P_{\phi}(k) | \Delta_l^{T,E}(k,\eta_0) |^2.
\label{eq:cldef}
\end{equation}
$P_{\phi}(k)$ is the initial power spectrum of scalar perturbations,
which are found not to give rise to any $B$ mode, ($C_l^B = 0$). The
transfer functions
\begin{equation}
\Delta_l^{T,E}(k,\eta_0) = \int_0^{\eta_0} d\eta \; S_{T,E}(k,\eta) f_l^{T,E} j_l(k[\eta_0 - \eta])
\label{eq:tfdef}
\end{equation}
are characterised by the source functions
\[
S_T(k,\eta) = e^{-\tau}(\dot{\xi}+\ddot{\alpha}) + \Lambda \biggl(
\Delta_{T,0} + \ddot{\alpha}\biggr) + \Lambda \biggl( \ddot{\alpha} +
\frac{\dot{v_b}}{k}\biggr)
\]
\begin{equation}
\phantom{xxxxxx}
+ \dot{\Lambda} \frac{v_b}{k} + \Lambda\biggl( \frac{\Pi}{4} + \frac{3\ddot{\Pi}}{4k^2}\biggr) + \dot{\Lambda}\frac{3\dot{\Pi}}{4k^2} + \ddot{\Lambda}\frac{3\Pi}{4k^2}
\label{eq:srcTdef}
\end{equation}
for the temperature and 
\begin{equation}
S_E(k,\eta) =  \frac{3\Lambda \Pi}{4(k[\eta_0-\eta])^2}
\label{eq:srcEdef}
\end{equation}
for the E-mode polarisation field. $f_l^{T} = 1 \; \forall l$ and $f_l^E =
\sqrt{(l+2)!/(l-2)!}$. In these equations, $j_l(x)$ are spherical Bessel
functions of order $l$, $\eta_0$ is the conformal time at the present epoch, and
$\alpha\equiv (\dot{\xi}+6\dot{\eta})/2k^2$ is a function of the metric scalar
perturbations $h$ and $\xi$ in the synchronous gauge, used throughout this
paper. The polarisation term $\Pi$, as already defined above, is the sum of the
temperature quadrupole and Q monopole and quadrupole, i.e. $\Pi \equiv
\Delta_{T,2} + \Delta_{P,2} + \Delta_{P,0}$, and $\Lambda = \dot{\tau}
e^{-\tau}$ is the visibility function.  $\dot{\tau}$ is the opacity, which, in
case of Thomson scattering, reads
\begin{equation}
\dot{\tau}_T = a\sigma_Tn_e,
\label{eq:thomsonopac}
\end{equation}
with $a$ the scale factor, $\sigma_T$ the Thomson cross section and
$n_e$ the comoving free electron number density. The optical depth is
defined as
\begin{equation}
\tau_T(\eta) = \int_{\eta}^{\eta_0} \dot{\tau}_T (\eta') d\eta'.
\label{eq:Thomsonoptdep}
\end{equation}
For these quantities, the subscript $T$ denotes "Thomson scattering".
Throughout this paper, dots denote derivatives with respect to conformal
time $\eta$.

\subsection{The Modified Visibility Function}
\label{sec:modvis}

We combine the previous results 
in order to compute the effect of resonant scattering on CMB
anisotropies. In equation (\ref{eq:rseq2}) we have seen that a resonant
transition of optical depth $\Delta \tau_{X}$ erases both I and Q
parameters by an amount $\Delta \tau_{rs}$, and generates new
intensity by an amount proportional to $\Delta \tau_{X}$ as well.
However, the new polarisation generated in the resonant scattering is
not proportional to $\Delta \tau_{X}$, but to $E_1 \; \Delta
\tau_{X}$. This fact requires the following modifications of
eqs.(\ref{eq:srcTdef},\ref{eq:srcEdef}) by means of the new visibility
function $\tilde{\Lambda}$:
\[
S_T(k,\eta) = e^{-\tau}(\dot{\xi}+\ddot{\alpha}) + \Lambda \biggl(
\Delta_{T,0} + \ddot{\alpha}\biggr) + \Lambda \biggl( \ddot{\alpha} +
\frac{\dot{v_b}}{k}\biggr)
\]
\begin{equation}
\phantom{xxxxxx}
+ \dot{\Lambda} \frac{v_b}{k} + \tilde{\Lambda}\biggl( \frac{\Pi}{4} + \frac{3\ddot{\Pi}}{4k^2}\biggr) + \dot{\tilde{\Lambda}}\frac{3\dot{\Pi}}{4k^2} + \ddot{\tilde{\Lambda}}\frac{3\Pi}{4k^2};
\label{eq:srcTdef2}
\end{equation}
\begin{equation}
S_E(k,\eta) = \frac{3\tilde{\Lambda} \Pi}{4(k[\eta_0-\eta])^2}.
\label{eq:srcEdef2}
\end{equation}
In this case, the total opacity and optical depth read
\begin{eqnarray}
\dot{\tau}(\eta) & = & \dot{\tau}_T (\eta) \; + \; \dot{\tau}_{X} (\eta), \\
\tau (\eta) & = & \int_{\eta}^{\eta_0} (\dot{\tau}_T + \dot{\tau}_{X}) d\eta'.
\label{eq:newtotaltau}
\end{eqnarray}
The visibility function $\Lambda$ remains unchanged in terms of
$\tau$ and $\dot{\tau}$, while the new visibility function
$\tilde{\Lambda}$ is defined as
\begin{equation}
\tilde{\Lambda} \equiv \bigl( \dot{\tau}_T + E_1 \dot{\tau}_{X}\bigr)\; e^{-\tau}.
\label{eq:vistilde}
\end{equation}
Note that this new visibility function is coupled to the polarisation terms
exclusively. The generation of anisotropies is ruled by the opacity term,
which does contain $E_1$, whereas the blurring of existing anisotropies is
caused by the exponential of $\tau$, which does {\em not} depend on $E_1$.
Hence, we expect that different resonant transitions with different
$E_1$ coefficients should be discernible in those angular scales ($l$ ranges)
where the generation of new polarisation anisotropies dominates or is 
comparable to the blurring of old anisotropies. I.e., we shall recover
no information on $E_1$ in the high $l$ limit for which $\delta C_l^{TT,EE} \simeq -2\tau \; C_l^{TT,EE}$. 

\begin{figure}
\begin{center}
        \epsfxsize=7cm \epsfbox{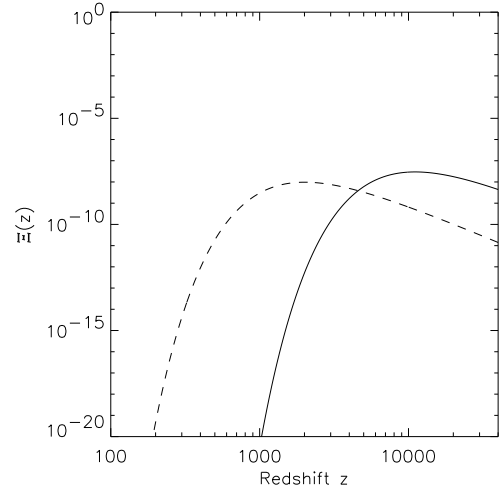}
\caption[fig:ratio]{
  Function $\Xi$ displaying the ratio of the drag force exerted by
  resonant scattering over the Thomson drag, for two different
  resonant transitions: Ly-$\alpha$ (solid line), and LiI
  2S$\rightarrow$ 2P [6708 \AA] (dashed line). At all redshifts, $\Xi$
  is much smaller than one, making the drag force induced by resonant
  species negligible if compared to the Thomson drag. }
\label{fig:ratio}
\end{center}
\end{figure}

\subsection{The Drag Force}

The Thomson scattering influences the evolution of baryons, since
they exchange momentum with the CMB photons. This
interaction is accounted for by adding a momentum conservation term to the
equation for the evolution of the baryon peculiar velocity, (e.g.,
\citet{mb}):
\begin{equation}
\dot{v}_b = -\frac{\dot{a}}{a} v_b - i c_s^2 k \delta_b + \frac{4\bar{\rho}_{\gamma}}{3\bar{\rho}_b}an_e\sigma_T (-\frac{3i}{4}F_{\gamma ,1} - v_b ).
\label{eq:vbaryon}
\end{equation}
$v_b$ denotes baryon peculiar velocity, $c_s$ is the sound speed in
the baryon photon fluid before recombination, $\delta_b$ is the
comoving baryon density contrast, and $\bar{\rho}_{\gamma (,b)}$ are
the photon (baryon) background energy densities. Finally, $F_{\gamma,
  1}$ is the $l=1$ moment in the expansion of the momentum averaged
photon energy distribution function, (see eq.(\ref{eq:fgammal})). When
considering the resonant scattering, a similar term should be included
in this equation. There is, however, a critical difference: Thomson
scattering is frequency independent, and couples {\em all} photons
with baryons, as opposed to resonant scattering, that couples only
those photons within the thermal width of the line.  \citet{loeb} uses
arguments based upon the characteristic time of the drag force to
conclude that it does not play any role in the case of Lithium
scattering. Here we explicitely compute the drag force, for both
Thomson scattering and resonant scattering, and confirm that indeed
the latter can be ignored in practically all cases.

For Thomson scattering, the drag force (defined here as rate of
momemtum exchange per unit time and unit volume) reads
\[
F_{\gamma}^T (\vk, \eta) = (-4\pi i)\int dq q^2 \sum_{i=1}^2 f_0^i(q,\eta) q\Psi_1^i(\vk, q,\eta)\; \times
\]
\begin{equation}
\phantom{xxxxxxxxxxxxxxxxxxxxxxxxxxxx}
 c\; \sigma_T n_e (\eta )
\label{eq:dgf_Thomson}
\end{equation}

but for resonant scattering on a transition $X$, the drag force must
acccount for limited range of frequencies that do interact with the baryons.
This is introduced via by imposing a resonance condition in
the momentum integration of the atom distribution function:
\[
F_{\gamma}^X (\vk, \eta) = (-4\pi i) \int dq q^2 \sum_{i=1}^2 f_0^i(q,\eta) q\Psi_1^i(\vk, q,\eta)\;\times
\]
\begin{equation}
\phantom{xxxxxxxxxxxxxxxxxxxxxx}
 c\;\sigma_X \int_{\Sigma}d\vp f_X(\vp,\eta) .
\label{eq:dgf_rt}
\end{equation}
The symbol $\Sigma$ denotes the subset of all possible atom momenta $\vp$
verifying the resonance condition.  $f_X(\vp,\eta)$ is the
distribution function of the atom responsible for the transition $X$.
For simplicity, we shall assume that all baryons are in thermal
equilibrium with the CMB radiation field at all redshifts, so that the
CMB temperature will characterise their Maxwell-Boltzmann distribution
function. The resonance condition
\begin{equation}
\frac{qc}{h} \biggl( \frac{\vp\cdot \n}{m_X c} - 1\biggr) = \nu_{X}
\label{eq:res_con}
\end{equation}
can be rewritten, after accounting for the natural width of the line, as
\[
\vp\cdot \n \in \biggl[ \mbox{min}\biggl( m_X c \left(1-\frac{\nu_X}{\nu} \pm \frac{A_X}{4\pi} \right)\biggr),
\]
\begin{equation}
\phantom{xxxxxxxxxxxx}
\mbox{max}\biggl(  m_X c \left(1-\frac{\nu_X}{\nu} \pm \frac{A_X}{4\pi} \right)\biggr)\biggr],
\label{eq:res_con2}
\end{equation}
$A_{X}$ is the Einstein coefficient for spontaneous emission of
transition $X$, and the ratio $A_{X}/2\pi$ is its natural width. $m_X$
and $\nu_X$ denote the mass of the resonant atomic species and the
resonant frequency, respectively. $\mu$ is the cosine of the angle between
$\vp$ and $\n$, ($\mu = \vp/p \cdot \n$). Since we are considering all
photon frequencies, all atoms within the thermal width are counted
and all of them contribute to the drag force, so the integral on
the atom distribution function yields $n_X(\eta )$, the atom number density,
(see Appendix A for details). Therefore:
\begin{equation}
F_{\gamma}^X \simeq (-4\pi i ) (c\;\sigma_Xn_X) \; \frac{hA_X}{2\pi
c}q_X^2 \;\left(f_0\Psi_1\right)_{q_X} \; q_X\;.
\label{eq:dgf_rt2}
\end{equation}
I.e., momentum of amplitude $q_X \equiv h\nu_X /c$ is transferred to the
atom fluid at a rate per unit volume $c\;\sigma_X n_X$ in a momentum-space
shell of radius $q_X$ and thickness $h\nu_X /(2\pi c)$.
The photon distribution momentum $f_0 \Psi_1$ corresponds to the sum over the
two polarisation states, $f_0\Psi_1 \equiv \sum_{i=1}^2 f_0^i\Psi_1^i$.
The ratio of the Thomson induced drag force and the one
generated by resonant scattering reads
\[
\frac{F_{\gamma}^X}{F_{\gamma}^T} \simeq \frac{\sigma_X n_X}{\sigma_T n_e}
\; \times
\]
\[
\phantom{xxxxxxx}
\biggl[  \frac{hA_X}{2\pi
c}q_X^2 \;\left(f_0(q_X,\eta)\Psi_1(\vk,q_X,\eta)\right) \; q_X\; \biggr]
\]
\begin{equation}
\phantom{xxxxxxxxxxxxxxxx}
\bigg/ 
\biggl[ \int dq\; q^2 f_0(q,\eta )\Psi_1(\vk,q,\eta) \; q  \biggr].
\label{eq:ratio_dfs}
\end{equation}
If we now consider that, at least {\it ab initio}, the $l=1$ momentum
of the photon distribution function $f_0(q,\eta) \Psi_1(\vq, q, \eta)$
is caused by gravitational sources, so that $\partial \Psi_1 /
\partial q \simeq 0$, then $\Psi_1$ is cancelled in
eq.(\ref{eq:ratio_dfs}), yielding
\[
\frac{F_{\gamma}^X}{F_{\gamma}^T} \simeq \frac{\sigma_X n_X}{\sigma_T n_e}
\biggl[  \frac{hA_X}{2\pi
c}q_X^2 \;f_0(q_X,\eta) \; q_X\; \biggr]\; \frac{4\pi c}{\bar{\rho}_{\gamma}}
\]
\begin{equation}
\phantom{xxxxxxxxxxxxxxxx}
\equiv
\frac{\sigma_X n_X}{\sigma_T n_e}\; \Xi[\eta, \nu_X].
\label{eq:ratio_dfs2}
\end{equation}
The ratio $\Xi $ versus redshift is shown in Figure (\ref{fig:ratio})
for two resonant transitions: Ly-$\alpha$ (solid line) and LiI
2S$\rightarrow$ 2P [6708 \AA] (dashed line). There are two factors
that make the drag force exerted by resonant transitions negligible
when compared to Thomson drag: {\it i)} Only photons within the line
width exchange momenta with the atoms, and {\it ii)} the line is, for
most redshifts, far from the peak of the planckian photon distribution
function. Therefore, for the moderate to small values of $\tau_{X}$
that we will be handling in this paper, the drag force can be
completely ignored in our calculations.

\begin{figure*}
\centering
\plotancho{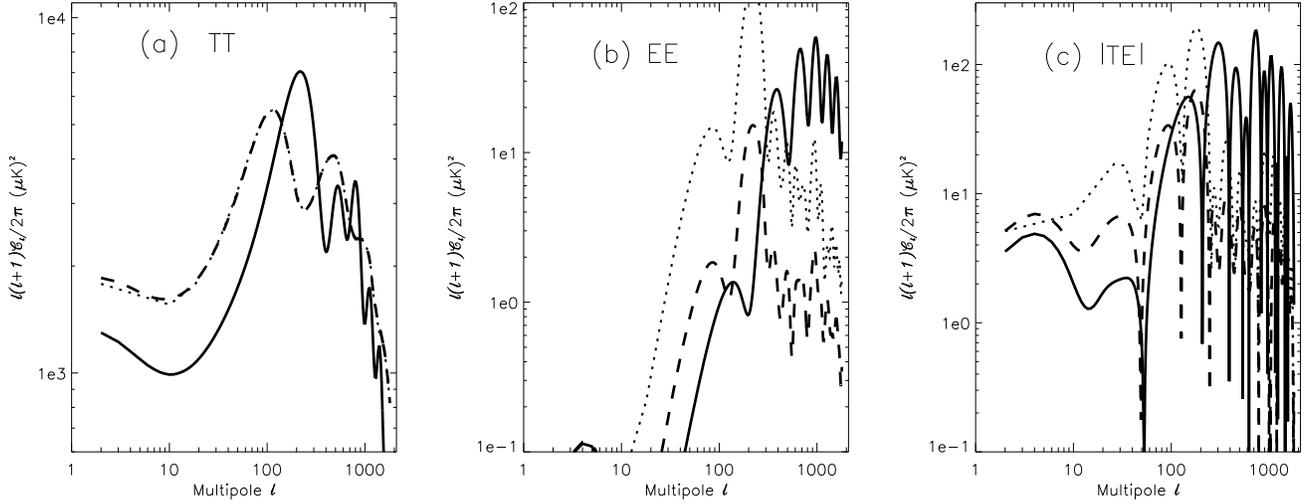}
\caption[fig:f_largetau]{TT, EE and TE angular power spectra for three different scenarios: standard $\Lambda$CDM cosmology (thick solid line), standard $\Lambda$CDM cosmology plus a resonant transition placed at $z_X \simeq 500$ and
$(\tau_{X}, E_1) = (2, 1/3)$ (dashed line), and standard $\Lambda$CDM cosmology plus a resonant transition placed at $z_X \simeq 500$ and
$(\tau_{X}, E_1) = (2, 1)$ (dotted line).
 }
\label{fig:f_largetau}
\end{figure*}
\begin{figure*}
\centering
\plotancho{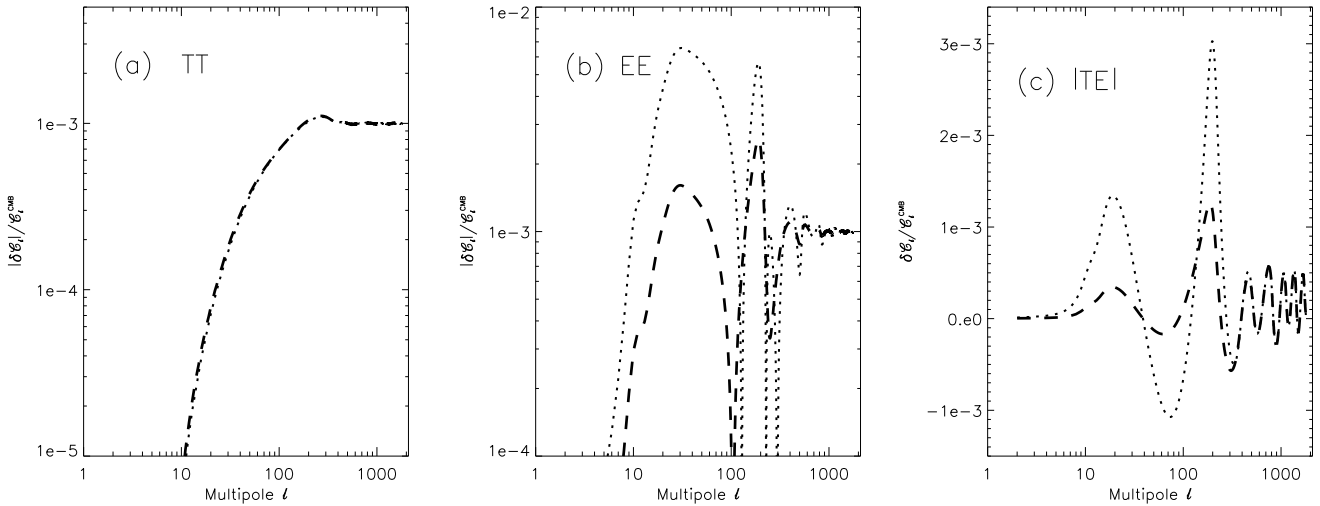}
\caption[fig:f_smalltau]{Relative increment of TT, EE and TE angular power spectra with respect to the standard $\Lambda$CDM scenario due to the presence of
  a resonant line placed at $z_X \simeq 500$ with $(\tau_{X},
  E_1)=(5\times 10^{-4}, 1/3)$ (dashed line) and $(\tau_{X},
  E_1)=(5\times 10^{-4}, 1)$ (dotted line). Note that for panels {\it
    (a)} and {\it (b)} we are plotting absolute values. Since the
  blurring of original anisotropies is independent of $E_1$, both
  lines converge to $2\tau_{X}$ in the high-$l$ range for the EE plot
  as well. Note that due to the change of sign of $C_l^{TE}$, in panel
  {\it (c)} we prefer to normalise by $\sqrt{C_l^{TT}C_l^{EE}}$.}
\label{fig:f_smalltau}
\end{figure*}
\subsection{Moderate and Low $\tau_X$ Examples}

In this subsection we compute the changes in the intensity and polarisation CMB
angular power spectra introduced by a resonant transition in two different
limits: moderate optical depth ($\tau_X = 2$), and small optical depth ($\tau_X
= 5\times 10^{-4}$). As shown in BHMS, in the limit of low $\tau_X$ there are
two competing terms: a generation term, present at scales larger than the
projected horizon at the epoch of scattering, and a blurring term, accounting
for the slight decrease in the amount of {\it previous-to-the-scattering}
temperature anisotropies. The latter is dominant at small angular scales, and
veryfies $\delta C_l^{TT} \simeq -2\tau_X \; C_l^{TT}$.  The former term is
caused by the growth of peculiar velocities with cosmic time, and is responsible
for positive $\delta C_l$'s below a multipole $l_{X}\sim (\eta_0 - \eta_X
) / \eta_0$, \citep{hms}.  In the limit of high $\tau_X$, however, the
generation of new anisotropies dominate in the entire angular range,
\citep{zaldaloeb}.

We consider a resonant transition at an effective scattering redshift close to
$z_X \simeq 500$, or conformal distance $\eta_X \simeq 450$ Mpc. This is close
to the epoch at which \citet{loeb,zaldaloeb} placed the resonant scattering
between the CMB and the neutral Lithium via the LiI 2S$\rightarrow$ 2P [6708
\AA] resonance. Although it has been shown that at such redshifts Lithium should
be ionised by Ly-$\alpha$ photons generated during recombination
\citep{hiratali}, we choose a similar scattering redshift for comparison
purposes.  We use a modified version of CMBFAST \citep{cmbfast} where a new
visibility function (modified as outlined in subsection (\ref{sec:modvis})) has
been introduced. Motivated by our results above, we also disconnect the drag
force exerted by the atoms. We choose a gaussian profile for the line opacity,
\begin{equation}
\dot{\tau}_X (\eta ) = \tau_X \frac{
    \exp{-\frac{\left( \eta - \eta_X\right)^2}{2\sigma_X^2}}}{\sqrt{2\pi\sigma_X^2}}.
\label{eq:opac_X1}
\end{equation}
The width of the line is taken to be one percent of $\eta_X$, but it would
ultimately depend on the spectral resolution of the detector.  Figure
(\ref{fig:f_largetau}) shows the temperature (TT), E-mode polarisation (EE) and
cross (TE) power spectra for three scenarios: the concordance $\Lambda$CDM model
(thick solid lines), used as a reference model, and this reference model as seen
after introducing a species giving rise to a resonant scattering of parameters
$[\tau_X$, $E_1]$) = $[2,1/3]$ (dashed lines) and $[\tau_X$, $E_1]$) = $[2,1]$,
(dotted lines). Panel {\it (a)} shows that different values of $E_1$ are not
discernible in the TT power spectrum, since the contribution of the polarisation
terms is of the order of a few percent: in the plot these differences are
visible to the eye at $l < 20$, but these are the scales more affected by Cosmic
Variance. Nevertheless, the EE power spectrum shown in panel {\it (b)} provides
very different estimates for different values of $E_1$, making both scenarios
easily discernible for this large value of $\tau_X$. Since new anisotropies are
generated in an older universe subtending a larger angle in the sky, the
angular pattern of the power spectra shiftes to lower multipoles, both in the TT
and the EE cases.
The polarisation source term $S_E (k, \eta)$ is larger at $z\sim 500$ than at
decoupling (see eq.(\ref{eq:srcEdef2})), and this makes the EE power
spectrum of the $[\tau_X$, $E_1]$) = $[2,1]$ of higher amplitude
(dotted line above solid lines). Note that the cross-power spectra
also show significant differences for the two $E_1$ values considered,
(Fig.(\ref{fig:f_largetau}c)).

The low $\tau_X$ case, displayed in Fig.(\ref{fig:f_smalltau}), also provides
significant differences between the $E_1=1$ and the $E_1=1/3$ models
in the TE and EE power spectra. These differences vanish in the
high $l$ limit, where the blurring term takes over and $\delta C_l \rightarrow
-2\tau_X\; C_l$ for all values of $E_1$. In case of resonant scattering on
metals produced at late epochs (end of the Dark Ages and reionisation),
the generation of new anisotropies arise at very large angular scales,
and hence distinguishing different values of $E_1$ will require large sky coverage.\\

\begin{table}
\caption{Possible $H_{\alpha}$ ($n=2 \rightarrow$  $n=3$) and $P_{\alpha}$ ($n=3 \rightarrow$ $n=4$) transitions$\;^{(a)}$}
\begin{tabular}{ccccc}
\hline
\hline
Wavelength &  $A_{ji}$ & $i$  & $j$  & $E_1$ \\
($\AA$) & ($s^{-1}$) & $\left( L_J\right) $& $\left( L_J \right)$ & \\
\hline
6562.7096  & 5.388e+07  &  $\;P_{1/2}$ & $\;D_{3/2}$ & 0.5 \\
6562.7247  & 2.245e+07  &  $\;S_{1/2}$ & $\;P_{3/2}$ & 0.5\\ 
6562.7517  & 2.104e+06  &  $\;P_{1/2}$ & $\;S_{1/2}$ & 0\\ 
6562.7714  & 2.245e+07  &  $\;S_{1/2}$ & $\;P_{1/2}$ & 0\\
6562.8516  & 6.465e+07  &  $\;P_{3/2}$ & $\;D_{5/2}$ & 0.28\\
6562.8672  & 1.078e+07  &  $\;P_{3/2}$ & $\;D_{3/2}$ & 0.32\\
6562.9093  & 4.209e+06  &  $\;P_{3/2}$ & $\;S_{1/2}$ & 0\\
\hline
18750.684 & 5.864e+06 & $\;P_{1/2}$ & $\;D_{3/2}$ & 0.50 \\
18750.720 & 3.065e+06 & $\;S_{1/2}$ & $\;P_{3/2}$ & 0.50\\
18750.829 & 6.117e+05 & $\;P_{1/2}$ & $\;S_{1/2}$ & 0\\
18750.881 & 3.065e+06 & $\;S_{1/2}$ & $\;P_{1/2}$ & 0\\
18751.011 & 7.037e+06 & $\;P_{3/2}$ & $\;D_{5/2}$ & 0.28\\
18751.011 & 1.287e+07 & $\;D_{3/2}$ & $\;F_{5/2}$ & 0.28\\
18751.064 & 3.475e+04 & $\;D_{3/2}$ & $\;P_{3/2}$ & 0.32\\
18751.065 & 1.173e+06 & $\;P_{3/2}$ & $\;D_{3/2}$ & 0.32\\
18751.111 & 1.379e+07 & $\;D_{5/2}$ & $\;F_{7/2}$ & 0.21\\
18751.138 & 9.193e+05 & $\;D_{5/2}$ & $\;F_{5/2}$ & 0.37\\
18751.191 & 3.128e+05 & $\;D_{5/2}$ & $\;P_{3/2}$ & 0.02\\
18751.210 & 1.223e+06 & $\;P_{3/2}$ & $\;S_{1/2}$ & 0\\
18751.225 & 3.475e+05 & $\;D_{3/2}$ & $\;P_{1/2}$ & 0\\
\hline\hline
\end{tabular}
\medskip

(a) Taken from NIST URL site: {\tt http://physics.nist.gov/PhysRefData/ASD/lines\_form.html}.\\
\label{tab:levels}
\end{table}

\begin{figure}
\begin{center}
        \epsfxsize=7cm \epsfbox{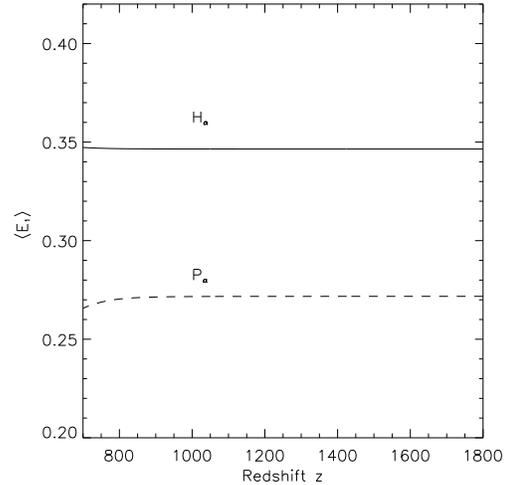}
\caption[fig:E1rec]{ Polarisation coefficients $E_1$ associated to $H_{\alpha}$ (solid line) and $P_{\alpha}$ (dashed line) during recombination. Among all transitions listed in Table (\ref{tab:levels}), only a few (one in case of $H_{\alpha}$, 3 in case of $P_{\alpha}$) contribute to the final value of $E_1$.}
\label{fig:E1rec}
\end{center}
\end{figure}

\section{Cosmological Hydrogen Recombination}

The first scenario where resonant scattering may modify the pattern of
the CMB polarisation anisotropies is precisely the cosmological
hydrogen recombination.
The CMB photons observed today were essentially last scattered during
the redshift range $z\in [700, 1300]$,, when the temperature of the
Universe was sufficiently low to permit the formation of neutral atoms
\citep{ZKS68,peebles68}.  During the process of recombination of the
hydrogen and helium atoms, there was a net generation of photons which
introduced distortions to the CMB blackbody spectrum. Except for the
far Wien tail of the CMB, these distortions are expected to be very
small \cite[e.g.][]{RCS06}, with typical values of the order of
$\Delta I/I \approx 10^{-7}$, making their detection challenging.
However, as studied in \citet{jalchmras}, these lines leave also an
imprint in the temperature angular power spectrum of the CMB. When
observing the CMB at a frequency in the vicinity of a redshifted
hydrogen line, the small optical depth of this transition influences
the angular distribution of the CMB, due to the resonant scattering of
the photons in this line. Therefore, we can use the optical depths
connected with these transitions to explore the consequences of the
overpopulation of the hydrogen atom levels at these epochs. \\

\begin{figure*}
\centering
\plotancho{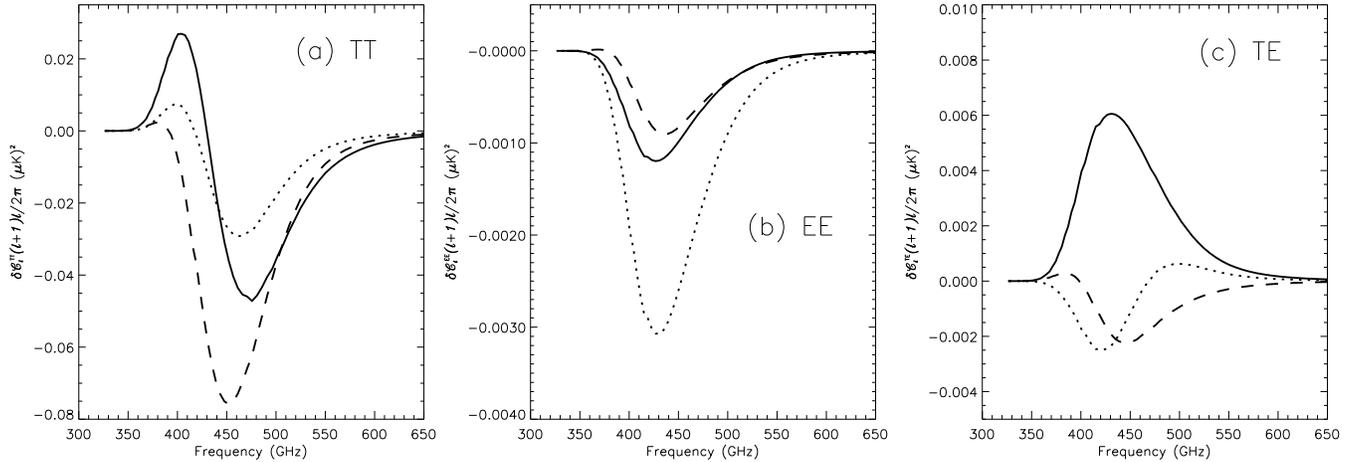}
\caption[fig:clsrec]{Change in the TT ({\it a} panel), EE ({\it b} panel) and TE ({\it c} panel) power spectra due to resonant scattering on $H_{\alpha}$ versus observing frequency. Solid, dashed and dotted lines correspond to multipoles $l=$ 783, 890 and 1001, respectively.} 
\label{fig:clsrec}
\end{figure*}

\begin{figure*}
\centering
\plotancho{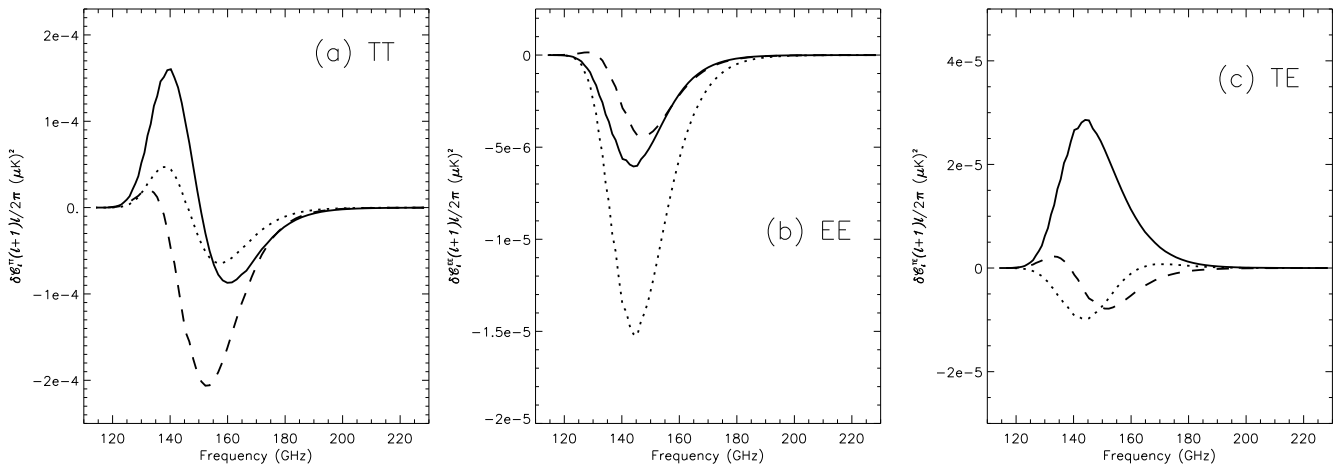}
\caption[fig:clsrec_Pa]{Same as in Figure (\ref{fig:clsrec}), but for the $P_{\alpha}$ line. Note the different amplitude and frequency ranges.
}
\label{fig:clsrec_Pa}
\end{figure*}

Here, we build upon the work of \citet{jalchmras} and compute the
change in the CMB polarisation anisotropies during hydrogen
recombination.  In particular we focus on the main ($\Delta n = 1$)
transitions of the Balmer and Paschen series, i.e. $H_{\alpha}$ and
$P_{\alpha}$: the first line provides the highest amplitude, 
whereas for the second line the corresponding redshifted frequency
lies in the frequency band where most of the future CMB experiments
will observe, and the foreground contamination is minimal.  Note that
in addition to the bound-bound transitions described in
\citet{jalchmras}, there is also a contribution from the free-bound
emission during recombination \citep{CS06}. However, in the considered
frequency band, their contribution to the total optical depth in the
redshift range where recombination occurs is very small \citep{CRS06}.

For the $H_{\alpha}$ line, there are seven permitted (electric
dipolar) transitions from the third shell ($n=3$) to the second one
($n=2$), depending on the particular configuration of the electronic
orbital, intrinsic (spin) and total angular momenta of the levels
involved. For the $P_{\alpha}$ case, the number of transitions amounts
to seventeen. In Table \ref{tab:levels} we list these transitions. For
each line, the total optical depth observed at the corresponding
frequency will be the sum of all contributing transitions. For each of
transition $j\rightarrow i$, the optical depth $\tau_{ij}$ is given by
\citep{sobolev}
\begin{equation}
\tau_{ij} = 
   \frac{A_{ji}\lambda_{ij}^3\left[n_i(g_j/g_i)-n_j \right]}{8\pi H(z)},
\label{eq:sob1}
\end{equation}
where $A_{ji}$ is the sponteaneous emission Einstein coefficient, $\lambda_{ij}$
is the wavelenght, and $n_{i-j}$, $g_{i-j}$ are the populations and degeneracy
factors of each of the two levels involved in the transition,
respectively. $H(z)$ is the Hubble function at redshift $z$. In our
computations, the populations for the different sublevels having different
orbital angular momenta ($2s, 2p, 3s, 3p, 3d, 4s, 4p, 4d$ and $4f$) were tracked
{\em independently}. Within each of these sublevels, we assummed that states
having different total angular momenta $j$ were {\em evenly} populated, i.e., as
it is dictated by the Clebsch-Gordan coefficients and/or the statistical
equilibrium condition.  This is a reasonable assumption given that the
energy differences between the different fine-structure sub-states of a given
shell are four order of magnitude smaller that its energy. Given that the
recombination lines are formed when the number of photons (per baryon) which are
able to excite them falls below one (see e.g. \citet{RCS06}), it is clear that
the fine-structure sub-levels will maintain equilibrium due to the ambient CMB
radiation field. Once we derive the optical depth for each transition, we
compute the average polarisation coefficient $E_1$ associated to both
$H_{\alpha}$ and $P_{\alpha}$ as
\begin{equation}
E_{1} = \frac{\sum_{ij} \tau_{ij} E_1^{ij}}{\sum_{ij} \tau_{ij}},
\label{eq:E1_Halpha}
\end{equation}
with $E_1^{ij}$ the polarisation coefficient for the particular
transition $i\rightarrow j$.  
Figure (\ref{fig:E1rec}) shows the behavior of both $E_1$ parameters
through the recombination epoch in these circumstances: they show a
somewhat constant value around $\sim 0.35$ for $H_{\alpha}$ and $\sim
0.27$ for $P_{\alpha}$. This is mostly due to the presence of few
dominant transitions (one for $H_{\alpha}$, three for $P_{\alpha}$)
having the largest weight in eq.(\ref{eq:E1_Halpha}). In the case of
$H_{\alpha}$, it is the $ P_{3/2} \rightarrow D_{5/2}$ transition
driving the final value of $E_1$ ($\sim 60$\% of the total $\tau$),
whereas for $P_{\alpha}$ these are $D_{7/2} \rightarrow F_{7/2}$
($\sim 40$\%), $D_{3/2} \rightarrow F_{5/2}$ ($\sim 25$\%) and $D_{5/2}
\rightarrow P_{3/2}$ ($\sim 7$\%).\\

Figure (\ref{fig:clsrec}) displays the changes in the TT, EE and TE
CMB angular power spectra versus observing frequency in the range
where the $H_{\alpha}$ is redshifted. The (remarkably smaller) amplitudes for
the $P_{\alpha}$ line are shown in figure (\ref{fig:clsrec_Pa}). We
have taken for the effective relative width of the line the value
$5\times 10^{-3}$, ($\sigma_X / \eta_X$, see eq.(\ref{eq:opac_X1})).
Solid, dashed and dotted lines correspond to multipoles $l=$ 783, 890
and 1001 (corresponding for the maximum change in TE, TT and EE power
spectra, respectively). These effects are considerably small, but
however might be within the sensitivity limit of upcoming CMB
experiments attempting to measure the B mode of polarisation
anisotropies. Actually, when looking at $H_{\alpha}$, the amplitudes
for the change in the TT and TE power spectra are comparable with the
expected amplitude of the B modes at intermediate angular scales ($l
\in [60,600]$) for a tensor to scalar ratio $r$ of 0.03. Instead, the
amplitudes for the changes in the EE power spectrum are closer to the
same prediction for the B mode at larger angular scales, ($l \in
[10,30]$). Also, the definite $l$ and frequency dependence of these
changes in the power spectra should make their detection more
tractable in practice, although at those frequencies dust is suppossed to
be dominating. In any case, we must conclude that, together with the
characterisation of the B modes of CMB polarisation anisotropies, the
detection of such small signals remains challenging for ongoing and
upcoming observational efforts.

\section{Reionisation}

In this Section, we revisit the imprint of resonant scattering of CMB
photons off metals produced at the end of the dark ages and the
beginning of reionisation. We shall attempt to build a somewhat
realistic model for the production of heavy elements during cosmic
times, and to obtain from it an amplitude for the effect of scattering
on metals produced by the first stars.

\subsection{Resonant Transitions}

A search for resonant transitions falling in a frequency range that
permits probing the resonant interaction with CMB photons in the
redshift range $z\in [5,50]$ yields the list shown in Table
\ref{tab:lines}. These are magnetic dipolar (M1) forbidden
transitions, associated to elements and ions that should be more
abundant during the epoch of cosmic dawn. After assuming a
(volume-average) metal abundance of one part in one hundred compared
to the solar value, we obtain typical values for the optical depth in
the range $\tau_X \in [10^{-4}, 10^{-6}]$. I.e., we shall be working
in the optically thin limit.  Note that for the OI 63.2 $\mu$m line
the polarisation coefficient is small ($E_{1} = 0.01$), since it will
have its repercusion in the forthcoming discussion.

\begin{table*}
\caption{Resonant transitions in atoms/ions expected to be produced during Reionisation}
\begin{tabular}{ccccccc}
\hline
\hline
Atom/Ion & $\lambda$  & Oscillator strength & Optical depth & 
Ionisation Potential &  Excitation Potential & $E_1$ \\
 & ($\mu$m) & $f_i$ & ($\nu = 250$ GHz, $X/X_{\odot}=10^{-2}$) & (eV) & (eV) & \\
\hline
   CI & 609.7 &  1.3$\times  10^{ -9}$ &  2.7$\times  10^{ -6}$ & 11.3 &  0.0 &  1.00 \\
   CI & 370.4 &  9.1$\times  10^{-10}$ &  2.7$\times  10^{ -6}$ & 11.3 &  0.0 &  0.35 \\
  CII & 157.7 &  1.7$\times  10^{ -9}$ &  7.9$\times  10^{ -6}$ & 24.4 & 11.3 &  0.50 \\
  NII & 121.9 &  2.8$\times  10^{ -9}$ &  4.5$\times  10^{ -6}$ & 29.6 & 14.5 &  0.35 \\
  NII & 205.2 &  3.9$\times  10^{ -9}$ &  4.9$\times  10^{ -6}$ & 29.6 & 14.5 &  1.00 \\
 NIII &  57.3 &  4.7$\times  10^{ -9}$ &  1.1$\times  10^{ -5}$ & 47.5 & 29.6 &  0.50 \\
   OI &  63.2 &  3.2$\times  10^{ -9}$ &  4.3$\times  10^{ -5}$ & 13.6 &  0.0 &  0.01 \\
   OI & 145.5 &  1.9$\times  10^{ -9}$ &  1.7$\times  10^{ -5}$ & 13.6 &  0.0 &  0.00 \\
 OIII &  51.8 &  6.5$\times  10^{ -9}$ &  9.6$\times  10^{ -5}$ & 54.9 & 35.1 &  0.35 \\
 OIII &  88.4 &  9.3$\times  10^{ -9}$ &  1.0$\times  10^{ -4}$ & 54.9 & 35.1 &  1.00 \\
  OIV &  25.8 &  1.0$\times  10^{ -8}$ &  2.2$\times  10^{ -4}$ & 77.4 & 54.9 &  0.50 \\
  SiI &  68.5 &  4.9$\times  10^{ -9}$ &  3.3$\times  10^{ -6}$ &  8.1 &  0.0 &  0.35 \\
  SiI & 129.7 &  6.2$\times  10^{ -9}$ &  3.0$\times  10^{ -6}$ &  8.1 &  0.0 &  1.00 \\
 SiII &  34.8 &  7.9$\times  10^{ -9}$ &  7.4$\times  10^{ -6}$ & 16.4 &  8.1 &  0.50 \\
   SI &  25.2 &  8.0$\times  10^{ -9}$ &  4.1$\times  10^{ -6}$ & 10.4 &  0.0 &  0.01 \\
   SI &  56.3 &  4.8$\times  10^{ -9}$ &  1.7$\times  10^{ -6}$ & 10.4 &  0.0 &  0.00 \\
 SIII &  18.7 &  1.8$\times  10^{ -8}$ &  1.1$\times  10^{ -5}$ & 34.8 & 23.3 &  0.35 \\
 SIII &  33.8 &  2.4$\times  10^{ -8}$ &  1.0$\times  10^{ -5}$ & 34.8 & 23.3 &  1.00 \\
  SIV &  10.5 &  2.6$\times  10^{ -8}$ &  2.0$\times  10^{ -5}$ & 47.2 & 34.8 &  0.50 \\
  FeI &  24.0 &  1.7$\times  10^{ -8}$ &  1.4$\times  10^{ -5}$ &  7.9 &  0.0 &  0.04 \\
  FeI &  34.7 &  2.1$\times  10^{ -8}$ &  1.4$\times  10^{ -5}$ &  7.9 &  0.0 &  0.03 \\
  FeI &  54.3 &  1.6$\times  10^{ -8}$ &  8.6$\times  10^{ -6}$ &  7.9 &  0.0 &  0.01 \\
 FeII &  26.0 &  1.7$\times  10^{ -8}$ &  1.4$\times  10^{ -5}$ & 16.2 &  7.9 &  0.05 \\
 FeII &  51.3 &  1.9$\times  10^{ -8}$ &  1.1$\times  10^{ -5}$ & 16.2 &  7.9 &  0.02 \\
 FeII &  87.4 &  1.1$\times  10^{ -8}$ &  4.6$\times  10^{ -6}$ & 16.2 &  7.9 &  0.00 \\
FeIII &  22.9 &  1.7$\times  10^{ -8}$ &  1.4$\times  10^{ -5}$ & 30.6 & 16.2 &  0.04 \\
FeIII & 105.4 &  7.8$\times  10^{ -9}$ &  3.0$\times  10^{ -6}$ & 30.6 & 16.2 &  0.00 \\
\hline\hline
\end{tabular}
\label{tab:lines}
\end{table*}

\begin{figure}
\begin{center}
        \epsfxsize=9cm \epsfbox{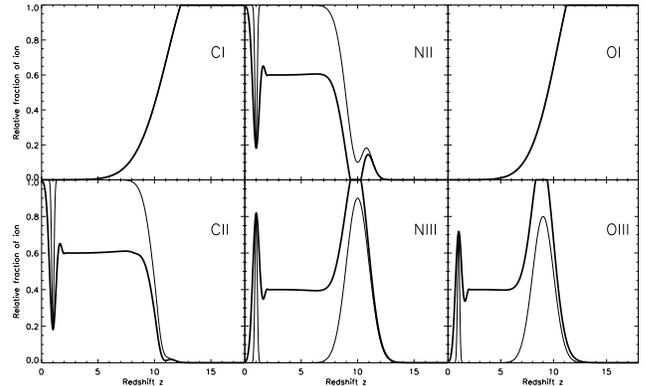}
\caption[fig:ion_species]{ Relative fraction of ions for two different
  reionisation scenarios: in case (A) Pop III stars at $z\sim 10$ are
  able to ionise three times C, N and O only temporarily, so that
  those ions are absent until a new starburst at around $z\sim 1$
  generates them again. This scenario is depicted by thin solid lines.
  In case (B), Pop III stars are able to keep a significant fraction
  of C, N and O ionised upto $z\sim 1$. In both cases, below $z\sim
  1$, the star formation rate drops considerably, and so do the
  abundances of highly ionised atoms.  }
\label{fig:ion_species}
\end{center}
\end{figure}

\begin{figure}
\begin{center}
        \epsfxsize=7cm \epsfbox{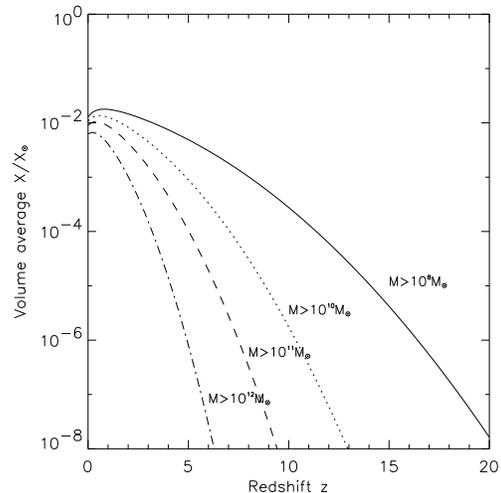}
\caption[fig:Xvol]{ Volume average metallicity measured by a CMB experiment
of angular resolution close to the arc-minute level and $\Delta \nu / \nu \sim 0.1$. }
\label{fig:Xvol}
\end{center}
\end{figure}

\subsection{A Toy Model for the Enrichment History of the Universe}

In a similar way to BHMS, we shall consider two different models of
reionisation: in case A we consider a reionisation scenario starting
at $z\sim 12$, that is able to ionise twice a significant fraction of
atoms like N, C and O {\em only temporarily} (down to redshift, say,
$z\sim 7$). Only after a new starburst takes place in galaxies at
$z\sim 1$ these ions appear again in the IGM. In case B, however, we
observe the scenario where double ionised species are present in the
IGM from $z\sim 10$ down to $z\sim 0.5$. The relative
abundances of ions for C, N and O for this toy model are given in Figure
(\ref{fig:ion_species}): thin lines correspond to case A scenario,
thick ones to case B. These elements correspond to the most abundance
species that are able to interact with CMB photons in the frequency
range of interest for us. Neutral Carbon and Hydrogen dissappear as
soon as first stars enter into scene for both reionisation models, but
singly ionised species (CII, OII, NII) will eventually re-appear only
if the flux of UV photons becomes lower than some threshold, (case A).
The pattern of the polarisation anisotropies will be particularly sensitive
to the presence or absence of certain species, as we next show.\\

\begin{figure*}
\centering
\plotancho{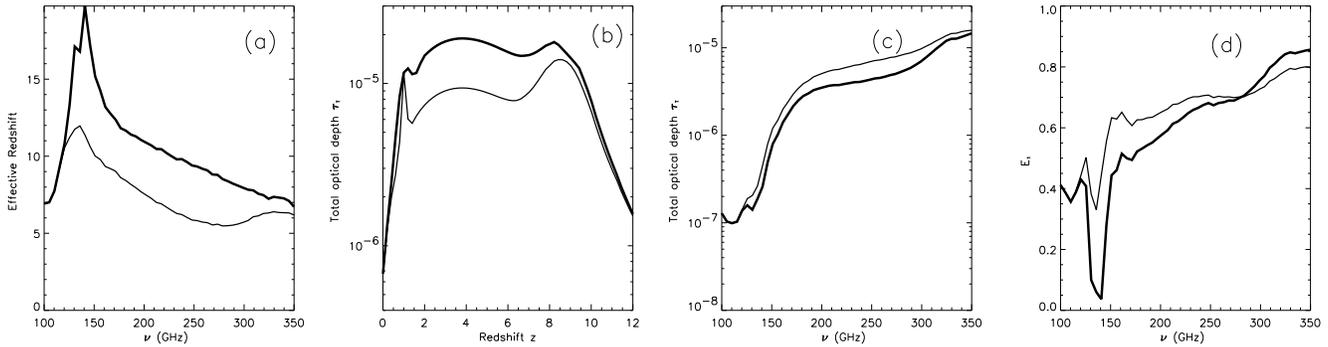}
\caption[fig:sumreio]{{\it (a)} Effective redshift versus observing frequency. {\it (b)} Total optical depth versus redshift {\it (c)} Total optical depth versus observing frequency. {\it (d)} Average polarisation coefficient versus observing frequency. In all panels, thin solid lines refer to case A reionisation scenario, thick solid lines to case B.} 
\label{fig:sumreio}
\end{figure*}

In our toy model, we shall assume that metals are located in the
interior of haloes more massive than a given threshold mass, here
taken to be $M_{th} \sim 10^8 M_{\odot}$. Inside these haloes the
metallicites are taken to reach the level of 5 times the solar
metallicity. In practice, due to the finite angular and spectral
resolution of the CMB experiment, future observations will be
sensitive to the {\em total } amount of atoms/ions within a given
cosmological volume probed by the CMB experiment, i.e., they will be
sensitive to the {\em volume average} metallicity of the IGM. In
Figure (\ref{fig:Xvol}) we show the volume average metallicity to be
measured by an experiment whose PSF and frequency response probes a
region comparable to the Hubble radius at each epoch, as it should be
the case for experiments having a spectral resolution of $\Delta \nu /
\nu \sim 0.1$. With this high dillution of the metallicities, the
resulting amplitudes for the resonant scattering will be accordingly
small.  For simplicity, we shall restrict our analyses to the first
ten lines considered in Table (\ref{tab:lines}), since they correspond
to the strongest line of the most abundant elements. The amplitudes and the
overall behaviour of the IGM metallicity given in Figure (\ref{fig:Xvol}) are not too different from observational estimates of \citet{schaye, aguirre, ledoux} or \citet{prochaska}
\\

\begin{figure}
\begin{center}
        \epsfxsize=7cm \epsfbox{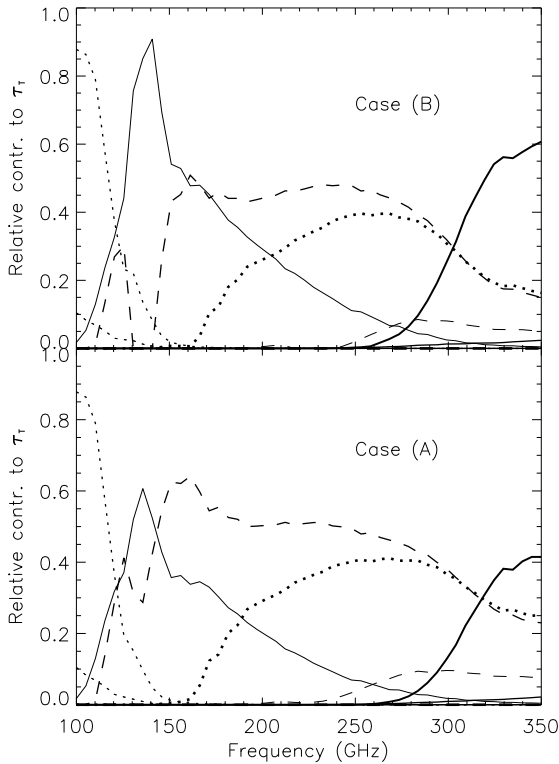}
\caption[fig:ratiotau]{Relative contribution of each line to the total optical depth versus observing frequency. Dotted, dashed and solid lines correspond to Carbon, Nitrogen and Oxygen, respectively. Increasing line thickness corresponds to increasing line of entry in Table \ref{tab:lines}. }
\label{fig:ratiotau}
\end{center}
\end{figure}

The fact that {\em all} existing lines will introduce some
anisotropies at redshifts dictated by the observing and resonant
frequencies translates into a non-trivial relation between observing
frequency and {\em effective redshift of scattering}. In Figure
(\ref{fig:sumreio}a) we display this effective scattering redshift
defined as
\begin{equation}
Z_{\mbox{eff}} (\nu ) \equiv 
        \frac{\sum_X z_X(\nu )\;\tau_X[z_X(\nu )]}{\sum_X \tau_X[z_X(\nu )] },
\label{eq:zeff_res}
\end{equation}
where the sum is over the resonant lines under consideration, and the
optical depth for each transition is considered to be a function of
the redshift corresponding to the observing frequency.  Again, the
thin solid line corresponds to case A, thick solid line to case B.
Although there is a general trend to have smaller effective redshifts
with increasing observing frequencies, the actual values may depend
critically on the dominant lines at each frequency. This is the cause
why panel {\it (b)} in Figure (\ref{fig:sumreio}) (total optical depth
$\tau_{Tot}$ versus redshift) differs so much from panel {\it (c)},
($\tau_{Tot}$ versus observing frequency): in panel {\it (b)} the
distinction between the case A and case B scenarios looks much more
feasible than in panel {\it (c)}. As shown in panel {\it (a)}, the
case B scenario is observing, on average, at higher effective
redshifts than case A, and this results in lower abundances and lower
optical depths, making it more similar to the case A scenario. Both
scenarios, case A and case B, could however be more easily
distinguished if measurements of the effective polarisation
coefficient were available; this coefficient, defined by
\begin{equation}
E_{1,\mbox{eff}} (\nu ) \equiv 
        \frac{\sum_X E_{1,X}\;\tau_X[z_X(\nu )]}{\sum_X \tau_X[z_X(\nu )] },
\label{eq:E1_eff}
\end{equation}
is plotted in Figure (\ref{fig:sumreio}d). The case B scenario (thick
line) shows a profound dip at $\nu \sim 140$ GHz, and its origin is
displayed in Figure (\ref{fig:ratiotau}). This last figure provides
the relative weight of each transition to the total optical depth at a
given observing frequency: dotted lines correspond to Carbon, dashed
lines to Nitrogen and solid lines to Oxygen. Increasing thickness
corresponds to increasing line of entry on Table \ref{tab:lines} for a
given element, from top to bottom. We see that neutral Carbon (thin
dotted lines) dominate the total optical depth at low frequencies,
i.e., high redshifts. On the contrary, at low redshifts (or high
frequencies) the optical depth is dominated by ionised Oxygen (thick
solid line). At intermediate redshifts, the lines from NII and CII
compete (medium thickness dotted and dashed lines, respectively). The
63.2 $\mu$m OI line (thin solid line) becomes prominent at $\nu \sim
140$ GHz, and in case B scenario where there is little NII (since most
of Nitrogen is in form of NIII) clearly dominates the total optical
depth in a narrow redshift range. Since the polarisation coefficient
associated to this transition is relatively low ($E_1 = 0.01$), it
leaves a distinctive signature in the $E_{1,\mbox{eff}}$. Of course, these
conclusions are consequences of our particular choice of reionisation
models, and are not generic.\\

Nevertheless, from a general point of view it is possible to state
that the OI 63.2 $\mu$m transition will always leave the signature in
the form of very low polarisation parameter $E_{1,\mbox{eff}}$, enabling its
isolation from the rest of resonant lines. Further, it is a prediction
for all models that the polarisation coefficient should increase with
increasing frequency, since in this regime it would be dominated by
the ionised species OIII, NII and CII. The measurement of polarisation
anisotropies generated during reionisation via CMB scattering off
metals would not only provide a consistency check for the detection in
the temperature/intensity side, but would also shed more light on how
reionisation took place. However, the detection of these polarisation
anisotropies would require sensitivities of the order of 1--100 $n$K,
and currently this is beyond the scope of most CMB polarisation
projects.

\section{Discussion and Conclusions}

The measurement and interpretation of the polarisation {\em intrinsic}
anisotropies of the CMB has remarkably tightened the constraints on
several cosmological parameters, \citep{wmap3}. However, their
characterisation is hampered by their low amplitude and the presence
of contaminating signals. To date only anisotropies of the E mode of
polarisation have been detected, but the detection of the B mode is
currently the main goal of many ambitious CMB projects. The main
difficulty that these projects will have to overcome is not only the
extremely demanding level of sensitivity required to measure such a
small signal, but also the exquisite control of systematics and
foregrounds contaminants, the latter being usually orders of magnitude above
the pursued component.\\

In this context, we have presented a study of the secondary
anisotropies of the CMB polarisation field caused via resonant
scattering in two different scenarios:
the cosmological recombination and the cosmological reionisation. 
In both cases, the new polarisation anisotropies are E type, but of
smaller amplitude than the intrinsic E type CMB polarisation
fluctuations. In the recombination scenario, we have focused our
analyses on the $H_{\alpha}$ line, giving rise to the largest signal
among all other transitions. We have found a definite and distinctive
pattern in multipole and frequency space for the change of the TE and
EE power spectra. This effect takes place in the frequency range $\nu
\in [350, 600]$ GHz, and is maximum at multipoles $l \sim 900$.  The
amplitudes of these signals are in general comparable to those of the
intrinsic B-mode polarisation anisotropies.
In a scenario where CMB experiments have the
sensitivity and systematics control level required to characterise the
B-mode anisotropies, the access to the study of the fluctuations
introduced by the $H_{\alpha}$ line should be then reachable.\\

The polarisation anisotropies introduced by resonant scattering off
metals during reionisation would provide a concluding consistency
check for the measurements carried out in the intensity/temperature
side. At the same time, they would shed some light on {\em which}
transitions are responsible for most of the measured effect. In
particular, the OI 63.2 $\mu$m line tends to lower the average
polarisation coefficient $E_1$ generated by all species, since it
conserves very little memory of the polarisation of the scattered
photon ($E_{1,63.2\mu m} = 0.01$). On the other hand, the dominant
transitions associated to highly ionised species, like CIII and OIII,
show relatively high values of $E_1$, ($E_1 \in [0.5, 1]$). The
requirements for detecting these signals are, however, even more
demanding than in the recombination scenario: typical amplitudes are
closer to the $n$K level than to the $\mu$K level, even below the
amplitudes expected for the primordial B-mode polarisation anisotropies. The
final detectability of this signal will be critically associated with
the detectability of the primordial B-mode anisotropies: if the
latter, at the end of the day, can be characterised with some given
accuracy, then the possibility of pursuing signals that are still 5 to
50 times smaller could be formally addressed.

\section*{Acknowledgments} 
CHM acknowledges fruitful conversations with R.Jimenez, Z.Haiman.
J.Garc\1a--Bellido and thanks L.Verde for carefully reading the
manuscript.  CHM is supported by NASA grants ADP04-0093 and
NNG05GG01G, and by NSF grant PIRE-0507768.  JARM acknowledges the
hospitality of the Department of Physics and Astronomy of the
University of Pennsylvania during his visit in June 2006.

\appendix

\section{The Drag Force}

A net flow of CMB photons will exert some drag force on a fluid of atoms that
are able to interact with the photon gas via some resonant transition $X$, with
drag force defined as "{\it rate of momentum exchange per unit time and unit
volumen}".  Here we detail the calculation of the drag force exerted on the atom
fluid for a given frequency of the CMB field. The expression will be integrated
in atom momentum and photon frequency to yield the well-known result that the
drag force will be proportional to $f_0(\nu_X,\eta)\Psi_1 \nu_X^3 n_X(\eta)$,
with $n_X(\eta )$ the average comoving number density of atoms taking part in
the resonance.

For a given transition X, the optical depth for scattering at a given frequency
is given by
\begin{equation}
\sigma (\nu) = \frac{c^2}{8\pi \nu_X^2} \frac{g_u}{g_l} A_{X} \Phi(\nu)
\frac{ 1 - \exp(-\frac{h \nu_X}{k T_{ex}}) }{1+ \frac{g_u}{g_l}\exp(-\frac{h
    \nu_X}{k T_{ex}})}
\label{eq:sigma_app}
\end{equation}
where $A_X$ is the Einstein coefficient for spontaneous emission of the
transition, $g_u$ and $g_l$ the statistical weights for the upper and lower
levels of the transition, respectively, and $T_{ex}$ represents the excitation
temperature, which in case of equilibrium with the radiation field would
correspond to the CMB temperature.  Finally, $\Phi(\nu)$ represents the line
profile. 

The photon flow and drag force are taken to be parallel to the polar axis in the
integration on $\vq$, (i.e., $\n$).  Only the dipole term
$f_0(q,\eta)\Psi_1(\vk,q,\eta)$ will contribute in the momentum integration, and
hence the drag force amplitude, expressed in terms of unit volumen occupied by
the atom fluid, can be written as
\[
F_{\gamma}^X = (-4\pi i)\int dq q^2 f_0(q,\eta)\Psi_1(\vk,q,\eta)\;q\; 
c\sigma_X\; 
\]
\begin{equation}
\phantom{xxxxxxxxxxxxxxxxxxxxxxx}  \;\int_{\Sigma} d\vp \; f_X(\vp, \eta).
\label{eq:df_ap1}
\end{equation}
For simplicity, we have introduced a frequency-independent cross-section
$\sigma_X$, which in practice is equivalent to approximate the profile function
by a top-hat function of a certain width $\Delta \nu$. Thus, the expression for
$\sigma_X$ is given by $\sigma(\nu)/ (\Delta \nu \Phi(\nu) )$. 
Note that $f_0\Psi_1 \equiv \sum_{j=1}^2 f_0^j\Psi_1^j$ corresponds to the
sum over polarisation states of the first momentum of the photon distribution
function expansion introduced in eq.(\ref{eq:phdf1}). The atom fluid is 
assumed to be described by a Maxwell-Boltzmann distribution function ($f_X$)
in thermal equilibrium with the CMB photon field:
\begin{equation}
f_X(\vp, \eta ) = \frac{n_X(\eta)}{\left[ 2\pi m_X k_B T\right]^{3/2}}
 \exp{\left( -p^2 / \left[ 2\pi m_X k_B T \right]\right) },
\label{eq:fX}
\end{equation}
with $T = T_0(1+z[\eta])$ the CMB temperature monopole.  The dominium
of integration $\Sigma$ is defined by the subset of possible values of 
$p$ and $\mu$ ($d\vp = p^2dpd\mu d\phi$) verifying the resonance
condition
\begin{equation}
\vp\cdot \n = m_X c \left( 1 - \frac{\nu_X}{\nu} \right),
\label{eq:rscond_ap}
\end{equation}
that is,
\[
\vp\cdot \n \in \biggl[ \mbox{min}\biggl( m_X c \left(1-\frac{\nu_X}{\nu} \pm \frac{A_X}{4\pi} \right)\biggr),
\] 
\begin{equation}
\phantom{xxxxxxxxxxxx}
\mbox{max}\biggl(  m_X c \left(1-\frac{\nu_X}{\nu} \pm \frac{A_X}{4\pi} \right)\biggr)\biggr],
\label{eq:rscond2_ap}
\end{equation}
with $A_X/2\pi$ the intrinsic natural width of the line.  Since the
integrand does not explicitely depend on $\mu$, we can integrate on
this variable, replacing $d\mu$ by $\Delta \mu = m_X c\; A_X /
(2\pi \nu p)$. Note that $\Delta \mu \leq 2$, and so
\begin{equation}
p \geq \frac{m_Xc}{2}\frac{A_X}{2\pi\nu}.
\label{eq:pmin1}
\end{equation}
Note, as well, that from the resonance condition of eq.(\ref{eq:rscond_ap})
we must have
\begin{eqnarray}
p & \geq & m_X c\biggl( 1 - \frac{\nu_X}{\nu}\biggr), \;\; (\nu > \nu_X), \\
p & \geq & m_X c\biggl( \frac{\nu_X}{\nu} - 1\biggr), \;\; (\nu < \nu_X).
\label{eq:pmin2}
\end{eqnarray}
\vspace{.2cm}
Therefore, the minimum value of $p$ allowed in the integration is
\begin{equation}
\pmin (\nu )= \mbox{max} \biggl[  \frac{m_Xc}{2}\frac{A_X}{2\pi\nu}, 
    \; \pm\; m_X c\biggl( 1 -\frac{\nu_X}{\nu}\biggr)\biggr].
\label{eq:pmin3}
\end{equation}
The first value will be used only if $\left| \nu_X / \nu -1
\right|\leq A_X / (4\pi \nu_X )$, and will give a negligible
contribution, as we shall see. 
The maximum value for $p$ will be
$\pmax = m_X c$. However, it will be approximated by infinity in the
integral, since $m_X c \gg \sqrt{2m_Xk_BT}$ in our case.  After
introducing $\Delta \mu$ in eq.(\ref{eq:df_ap1}), trivially
integrating in $d\phi$, and integrating in $p$ one obtains
\[
F_{\gamma}^X = (-4\pi i ) \int dq q^2 f_0\Psi_1\; q \;c\sigma_X n_X\;
   \frac{A_X}{2\pi \nu}\frac{m_X c}{\sqrt{2 m_X k_B T}}
\]
\begin{equation}
\phantom{xxxxxxxxxxxx}
   \;\;\exp{\left(-\pmin^2 (\nu ) / (2mk_B T)\right)}.
\label{eq:df_ap2}
\end{equation}
This equation shows that, for a given frequency, the
fraction of atoms being affected by the drag of CMB photons is
\begin{equation}
r_{X} = \frac{A_X}{2\pi \nu}\frac{m_X c}{\sqrt{2 m_X k_B T}}
\;\;\exp{\left(-\pmin^2 (\nu ) / (2mk_B T)\right)}.
\label{eq:ratio_atoms}
\end{equation}
>From eq.(\ref{eq:pmin3}), it is clear that the largest contribution of
the integral in eq.(\ref{eq:df_ap2}) comes from the vicinity of $q=q_X
= h\nu_X/c$, and this allows rewriting most of the terms out of the
$q$ integral, yielding
\[
F_{\gamma}^X \simeq (-4\pi i ) q_X^2 \;(f_0\Psi_1)_{q_X} \; q_X\; (c\;
\sigma_Xn_X) \; \frac{A_X}{2\pi \nu_X}\frac{m_X c}{\sqrt{2 m_X k_B
T}}\;
\]
\begin{equation}
\phantom{xxxxxxxx}
\int_{\mbox{line}} dq \; \exp{\biggl( -\pmin^2(\nu ) / (2m_Xk_BT)\biggr)}
\label{eq:df_ap3}
\end{equation}
We next sweep the vicinity of the line, and
split this integral in three separate ones, according to the three
different possible values of $\pmin (\nu )$. The final result is:
\[
F_{\gamma}^X \simeq (-4\pi i ) (c\;\sigma_Xn_X) \; \frac{hA_X}{2\pi
c}q_X^2 \;(f_0\Psi_1)_{q_X} \; q_X\;
\]
\[
\phantom{xxxxxxxx}
\biggl[ \mbox{erfc}\left(\frac{A_X}{4\pi\nu_X}\right) \; +
\]
\begin{equation}
 \phantom{x}
\frac{mc}{\sqrt{2\pi m_X k_BT}}\frac{A_X}{2\pi\nu_X}
\exp{-\left(\frac{m_Xc}{\sqrt{2m_Xk_BT}} \frac{A_X}{4\pi\nu_X}\right)^2}
\biggr].
\label{eq:df_ap4}
\end{equation}
The first term in brackets accounts for the integral throughout the
thermal width of the line. The argument of the complementary error
function is very close to zero, so this term is very close to unity.
That is, practically {\em all} atoms are experiencing some drag force
exerted by the CMB photons. The second term in brackets provides the
integral within the natural width of the line, and its contribution
can be safely neglected. Hence
\begin{equation}
F_{\gamma}^X \simeq (-4\pi i ) (c\;\sigma_Xn_X) \; \frac{hA_X}{2\pi
c}q_X^2 \;(f_0\Psi_1)_{q_X} \; q_X\;.
\label{eq:df_ap5}
\end{equation}
I.e., the momentum $q_X = h\nu_X / c$ is transferred to {\em all}
atoms at a rate given by $c\;\sigma_Xn_X$ in a momentum shell of
radius $q_X$ and thickness $hA_X/(2\pi c)$.

\label{lastpage}

\end{document}